\documentclass[prl,twocolumn]{revtex4}
\usepackage{graphicx}

\def\eqa{\begin{eqnarray}}
\def\eea{\end{eqnarray}}
\newcommand{\eq}{\begin{equation}}
\newcommand{\ee}{\end{equation}}

\begin{document}

\title{Competing orders and inter-layer tunnelling in cuprate superconductors:
A finite temperature Landau theory}
\author{Jian-Bao Wu, Ming-Xu Pei, and Qiang-Hua Wang}
\address{National Laboratory of Solid State
Microstructures,Institute for Solid State Physics, Nanjing
University, Nanjing 210093, China}


\begin{abstract}
We propose a finite temperature Landau theory that describes
competing orders and interlayer tunneling in cuprate
superconductors as an important extension to a corresponding
theory at zero temperature [Nature {\bf 428}, 53 (2004)], where
the superconducting transition temperature $T_c$ is defined in
three possible ways as a function of the zero temperature order
parameter. For given parameters, our theory determines $T_c$
without any ambiguity. In mono- and double-layer systems we
discuss the relation between zero temperature order parameter and
the associated transition temperature in the presence of competing
orders, and draw a connection to the puzzling experimental fact
that the pseudo-gap temperature is much higher than the
corresponding energy scale near optimum doping. Applying the
theory to multi-layer systems, we calculate the layer-number
dependence of $T_c$. In a reasonable parameter space the result
turns out to be in agreement with experiments.
\end{abstract}

 \maketitle

The essential phenomenology of high-$T_c$ superconductors are the
dome-shaped superconducting phase diagram as a function of doping
$x$, and the existence of a pseudo-gap normal state. The mechanism
of the novel superconductivity remains to be a challenge to
researchers. The fact that superconductivity follows from doping
anti-ferromagnetic Mott insulators and that the transition
temperature depends on doping non-monotonically suggest the
importance of the effects of strong correlation among the
electrons, as emphasized firstly by Anderson. \cite{anderson} A
theory in this line must go beyond the usual
Bardeen-Cooper-Schrieffer mean field theory. However, there is
also a possibility that the phenomenon is not that exotic, and a
modification to the conventional theory by including a competing
order may do the job. This is the attitude taken by Chakravarty,
Kee and V\"{o}lker \cite{ch}, who proposed a zero temperature
Landau theory to explain the general phase diagram, and in
particular the copper-oxide layer-number $N$ (within a unit cell)
dependence of superconducting transition temperature $T_c$ of
homologous series of cuprate superconductors.\cite{TcN}
%
%
The new ingredients are just a competing order, {\em e.g.}, the
d-density-wave (DDW),\cite{ddw} and inter-layer tunnelling that
enhances superconductivity. \cite{interlayer} It is therefore
important to judge how robust the conclusions are against the weak
points in the theory. Indeed, a few of them are debatable even
within the mean field framework itself. 1) When two orders compete
with each other, it is not clear whether there is still a definite
connection between the zero temperature order parameter and the
transition temperature even if this is the case in the absence of
competition between the orders. 2) In the case of multi-layer
systems, it is not clear which representative of the modulated
order parameters (due to charge redistribution) is most
appropriate to be related to the transition temperature, even if
one assumes that the zero temperature order parameter scales with
the transition temperature. 3) Moreover, the underlying motivation
for a competing order to superconductivity is the fact that the
normal state pseudo-gap seems to be independent of the pairing
gap. From the elaborate collection of data in Ref.\cite{loram} the
pseudo-gap energy scale extrapolates to zero at a doping level of
$x=0.19$, while the pseudo-gap phenomena certainly exists at
$T\geq T_c$ even at the same doping. This already indicates that
the zero temperature value of the competing order, assumed to be
responsible for the pseudo-gap, does not scale with the transition
temperature for the pseudo-gap itself (if the pseudo-gap
temperature crossover is a phase transition at all). As a
compromise, the authors in Ref.\cite{ch} defined a gap that is the
root-sum-square of both order parameters, and used it to represent
the pseudo-gap temperature. In general, there is no microscopic
basis for this gap-combination. In a microscopic tight-binding
model, the two gaps do not combine this way once the DDW bands is
doped away from half filling. 4) Finally the weak points beyond
the mean field theory is of course the ignorance of quantum or
thermal fluctuation of the order parameters.

The last point can only be addressed by going beyond mean field
theory, as in a recent spin wave analysis of the anisotropic
XY-model,\cite{phase} which reproduces nicely the experimental
$T_c(N)$ in homologous series of HBCO superconductors.\cite{TcN}
In this paper, we address points 1)-3) listed above in the mean
field framework under the motivation that the importance of a
successful mean field theory should not be underestimated. In
order to do so, we propose a finite temperature Landau theory, and
determines the transition temperatures for both orders
unambiguously. The structure of the rest of the paper is as
follows. A finite temperature theory for the superconducting order
or the DDW order alone is first proposed. We then introduce the
coupling between these two orders, and discuss how competing
orders in the case of mono-layer and double-layer systems, where
no charge redistribution occurs, modify the relations between zero
temperature order parameter and the transition temperature. By
comparing the theoretical phase diagram with experiments we fix
the parameter introduced in the finite temperature theory. We also
comment on point 3) raised above. Finally we extend the theory to
the multi-layer systems, and calculate the $T_c(N)$. We find that
in a reasonable parameter space, the theoretical results are in
agreement with experiment,\cite{TcN} and lend a support to the
conjecture from the zero temperature Landau theory.\cite{ch}

Let us begin with the simple case that the superconducting order
is the only order parameter. It is known that $2\psi_0/T_c=3.52$
in the weak coupling s-wave superconductor, where $\psi_0$ is the
zero temperature energy gap. For the case of weak coupling d-wave
superconductor, this ratio is roughly $4.16$. A Landau-type theory
could be derived near the transition temperature from an effective
microscopic model, but is a daunting task to access the zero
temperature limit. We would therefore take a phenomenological
attitude. We demand that the theory is of the form near $T_c$, and
while extended to lower temperatures should give a qualitative
temperature dependence of the order parameter, and in particular,
should yield a prescribed ratio of $\psi_0/T_c$. Without loss of
generality, we re-scale the temperature so that $\psi_0/T_c=1$.
Under these conditions, a suitable finite temperature Landau free
energy density is as follows, \eqa f_s=[\alpha(x)+\beta
T^2]|\psi|^2+\frac{\beta}{2}|\psi|^4.\eea Henceforth we use
arbitrary units, and borrow the parameterization in Ref.\cite{ch},
with $\alpha(x)=10(x-0.3)$ and $\beta=2$. This simple model would
yield $T_c=\psi_0=\sqrt{-\alpha(x)/\beta}$. We note that near
$T_c$ the coefficient of $|\psi|^2$-term can be approximated as
$2\beta T_c(T-T_c)$, which is of the desired form in a usual
Landau theory.

The same consideration can be applied for the DDW order alone,
with the free energy density \eqa
f_D=[a(x)+qbT^2]D^2+\frac{b}{2}D^4,\eea where $a(x)=27(x-0.22)$,
$b=2$, and $q$ is a new phenomenological constant. The transition
temperature for the DDW order alone is
$T_D=\sqrt{-a(x)/qb}=D_0/\sqrt{q}$, where $D_0$ is the bare zero
temperature DDW order. Since we have taken the latitude to set
$\psi_0/T_c=1$, there is no further freedom to set $q=1$. In fact
the value of $q$ is given by $q=(D_0/T_D)^2/(\psi_0/T_c)^2$
independently of the temperature re-scaling. It is therefore
understood that a smaller value of $q$ means that the DDW order is
more tolerant to thermal suppression, and vice versa. We take $q$
as a new phenomenological parameter, which will be fixed by
arguments below.

Let us now couple the two order parameters to form a theory
describing mono-layer systems, \eqa f=& &[\alpha(x)+\beta
T^2]|\psi|^2+\frac{\beta}{2}|\psi|^4\nonumber\\
& &+[a(x)+qbT^2]D^2+\frac{b}{2}D^4 +gD^2|\psi|^2,\eea where
$g=1.2$ is the coupling constant. As has been illustrated in
Ref.\cite{ch}, this theory yields zero temperature order
parameters $\psi_0(x)$ and $D_0(x)$ that depends on the doping
level $x$ in very much the same way as the empirical $T_c(x)$ and
$T_D(x)$, provided that one assumes that $T_D\propto
\sqrt{\psi_0^2+D_0^2}$. Here $T_D$ represents the pseudo-gap
temperature scale. However, as has been pointed out, there is no
microscopic basis for the gap-combination, and the relation
between such a combined gap and the pseudo-gap temperature is not
clear {\it a priori}. In our case, the transition temperature
$T_c$ for $\psi$ is determined by \eqa \alpha(x)+\beta
T_c^2+gD^2=0,\eea where $D=\sqrt{-[a(x)+qbT_c^2]/b}$ is the value
of $D$ at $T=T_c$. Similarly, the transition temperature $T_D$ for
$D$ is given by \eqa a(x)+bqT_D^2+g\psi^2=0,\eea where
$\psi=\sqrt{-[\alpha(x)+\beta T_D^2]/\beta}$ is the value of
$\psi$ at $T=T_D$. Explicit expressions for $T_c$ and $T_D$ are
not provided because $D\geq 0$ and $\psi\geq 0$ have to be imposed
in the above equations. We present $T_c$ (thick solid lines) and
$T_D$ (thin solid lines) for typical cases, with $q=0.1,0.6,1,1.5$
in Figs.1(a)-(d) respectively. We have also shown $\psi_0$ (thick
dotted lines) and $D_0$ (thin dotted lines) in each figure for
comparison to the transition temperatures. We observe that
$\psi_0/T_c=1$ for all doping levels in Fig.1(b). However, this
does not hold in the other cases. Moreover, in the same case
$D_0/T_D$ is apparently not a constant. The reason why Fig.1(b) is
a special case is because the parameters satisfy $qb-g=0$, which
gives rise to a line of critical temperatures for DDW at the
zero-temperature critical doping level. For $q<0.6$, as in
Fig.1(a), we observe a lobe-shaped DDW phase transition line near
$x=0.198$, namely, two $T_D$'s at a given doping. This is best
understood in the limit of $q=0$. Under this assumption DDW order
is not suppressed by thermal effects but by the superconducting
order. Near $T_c$ the latter is small and DDW may survive.
However, with lowering temperature the superconducting order
increases and eventually squeeze the DDW order. With a finite but
small $q$ the qualitative behavior is the same. Because such a
feature is not observed yet, it is very unlikely that $q<0.6$. For
larger values of $q$ as in Figs.1(c) and (d), the resulting doping
dependence of $T_c$ is very oblique, a feature that does not seem
to appear in high-$T_c$ phenomenology. Basing on these judgements
we believe a reasonable regime is given by $0.6\leq q<1$. We
emphasize an interesting feature in Fig.1(b), where $T_D$ survives
above $T_c$ in the under-doped regime and drops abruptly to zero
at $x=0.198$, whereas $D_0$ vanishes nearby. The reason for this
to occur is because $D$ vanishes here not because of thermal
suppression but the competing order $\psi$. This seems to account
for the puzzling fact that even though pseudo-gap behavior is
observed at relatively high temperatures, but the associated
energy scale extrapolates to zero near $x=0.19$.\cite{loram} We
regard this as yet another successful aspect of this mean field
theory.

\begin{figure}
\includegraphics[width=8.5cm,height=7cm]{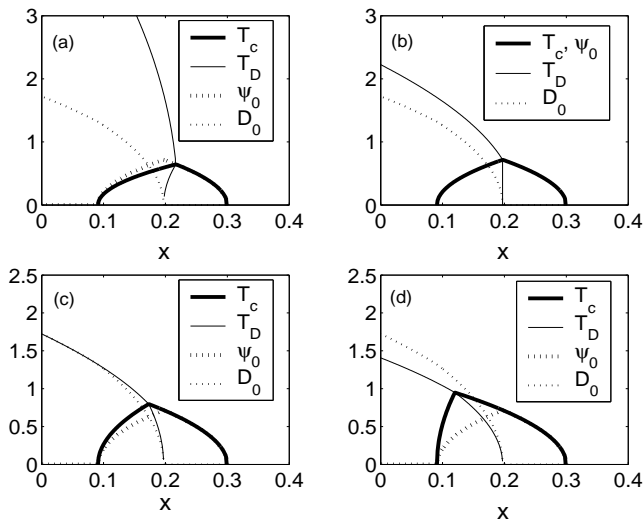}
\caption{Transition temperatures $T_c$ (thick solid lines) and
$T_D$ (thin solid lines), and zero temperature order parameters
$\psi_0$ (thick dotted lines) and $D_0$ (thin dotted lines) as
functions of doping level $x$ in a mono-layer system for (a)
$q=0.1$, (b) $q=0.6$, (c) $q=1$ and (d) $q=1.5$. Note that $T_c$
and $\psi_0$ coincides wherever $T_D=0$ according to the
convention used in the text, and they coincides in (b) at all
doping levels.}
\end{figure}

We now consider a double-layer system, where the only modification
to the theory is the inter-layer tunnelling, \eqa f=& &
\sum_{n=1}^2\left([\alpha(x)+\beta
T^2]|\psi_n|^2+\frac{\beta}{2}|\psi_n|^4\right.\nonumber\\ & &
\left. +[a(x)+qbT^2]D_n^2+\frac{b}{2}D_n^4 +gD_n^2|\psi_n|^2\right)\nonumber \\
& &-J(\psi_1^*\psi_2+\rm{c.c}),\eea where $\psi_n$ and $D_n$
represents the order parameters on the $n$-th layer, and $J=0.3$
is the tunnelling energy scale. By symmetry the equilibrium values
of the order parameters are independent of the layer index.
Therefore the analysis can be proceeded in very much the same way
as for the mono-layer case. The effect of the inter-layer
tunnelling can be included into a modification of the upper doping
limit of the superconducting order, so that
$\alpha(x)-J\rightarrow 10(x-0.33)$. The discussion then follows
closely the case of mono-layer. Figs.2 present the zero
temperature order parameters and the transition temperatures for
the corresponding values of $q$ used in Figs.1. Because of
interlayer tunnelling that enhances superconducting order, the
onset doping for the zero temperature DDW order is reduced
slightly to $x=0.188$. Except from such details, we find that the
qualitative behavior is the same as in a mono-layer. In particular
$bq-g=0$ remains to be the special case.

\begin{figure}
\includegraphics[width=8.5cm,height=7cm]{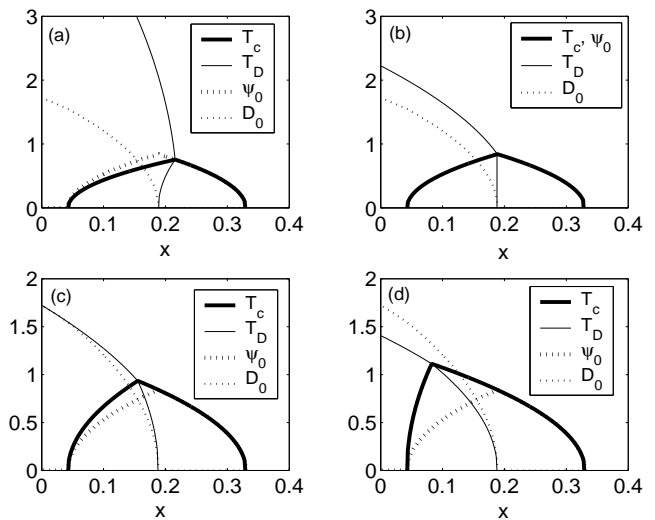}
\caption{The same plot as Figs.1, but for a double-layer system
with inter-layer tunnelling, because of which the upper doping
limit of the superconducting order shifts to $x=0.33$, and the
onset of zero-temperature DDW order occurs at $x=0.188$.}
\end{figure}

Given the success of the theory in dealing with mono- and
double-layer systems, generalization to more layers seems natural
and reasonable. The free energy density is now written as, \eqa
f=& &\sum_{n=1}^N\left([\alpha(x_n)+\beta
T^2]|\psi_n|^2+\frac{\beta}{2}
|\psi_n|^4\right.\nonumber\\ & &-J(\psi_n^*\psi_{n+1}+\rm{c.c})\nonumber\\
&&\left.+[a(x_n)+qbT^2]D_n^2+\frac{b}{2}D_n^4+gD_n^2|\psi_n|^2\right),
\eea where $N$ is the total number of layers, and $x_n$ is the
doping level on the $n$-th layer, which is different from the mean
doping level $x$ because of the charge redistribution
effect.\cite{charge} The hole distribution can be roughly
described by $x_n=[1-\epsilon/(N-2)]x$ for the inner layers
$1<n<N$ and $x_n=[1+\epsilon/2]x$ for the outer layers ($n=1$ and
$n=N$).\cite{charge,ch} Here $\epsilon=0.085$, 0.39 and 0.61 for
the $N=3,4,5$ respectively. We note that the hole distribution was
measured experimentally mainly in the doping regime $x\geq
0.18$.\cite{charge} Anticipating that $\epsilon$ does not change
appreciably with doping (for a fixed $N$), we shall extrapolate to
obtain the hole distribution in the under-doped regime.
With the present finite temperature theory, we can calculate $T_c$
directly from the secular problem in the linearized Landau
equations, \eqa & & [\alpha(x_n)+\beta T^2+gD_n^2]
\psi_n-J(\psi_{n+1}+\psi_{n-1})=0, \label{linear}\eea where $
D_n^2=\max\{0,-[a(x_n)+qbT^2]/b\}$ and $n=1,2,...,N$. The
condition for the existence of a nontrivial solution to the linear
homogeneous Eqs.(\ref{linear}) gives uniquely $T=T_c$. We seek
symmetric solutions for $\psi_n$ as a function of $n$, as this
yields the highest $T_c$. This reduces the number of independent
variables $\psi_n$, so that up to $N=5$ we only have to deal with
a $3\times 3$ matrix determinant. The algebra is straightforward.
The numerical result of $T_c$ as a function of $N$ is presented in
Figs.3 for the cases of (a) $x=0.14$ (under-doped), (b) $x=0.2$
(optimally doped) and (c) $x=0.25$ (over-doped). In each case we
used $q=0.6-1$, as argued to be reasonable previously. In Fig.3(a)
$T_c$ peaks broadly at $N=3$ for $q=1$. We have checked that for
even lower doping levels $T_c(N)$ increases monotonically (and
eventually saturates), to which we shall return shortly.
Homologous serious in this doping region is not available to the
best of our knowledge but is desirable to check the prediction. In
Fig.3(b) $T_c$ peaks at $N=3$ ($4$) for $q=0.6$ ($0.8$), and
explains nicely the data reported in Ref.\cite{TcN,charge}. In
Fig.3(c) $T_c$ increases monotonically again. This is also
consistent with experimental data. Indeed, $T_c(3)=77K$ and
$T_c(4)=117K$ at $x\approx 0.25$ in Table 1 of ref.\cite{charge}.
Superconductors with $N\geq 5$ and $x\geq 0.25$ are not available
in Ref.\cite{charge}.

The non-monotonic dependence in $T_c(N)$ is understood as a
cooperative effect from inter-layer tunnelling, charge
redistribution, and the competing order. For a uniform hole
distribution, the effect of inter-layer tunnelling is to enhance
superconductivity so that $T_c$ increases with the average
layer-coordination number $2-2/N$ (as has been checked but not
shown here). This accounts for the initial enhancement in $T_c(N)$
for all doping levels. In the very under-doped region the hole
density is low despite of the charge redistribution effect. In
such cases, the competing DDW order is robust and does not change
appreciably. Therefore, apart from a global suppression by DDW,
$T_c$ should increase with $N$. This is the case in Fig.3(a) if
one takes $q=0.6$ as the appropriate parameter. In the very
over-doped region, the hole density is so high that DDW order is
absent (at least up to $T_c$), and only the inter-layer tunnelling
matters, yielding an increasing $T_c(N)$ again, as in Fig.3(c).
The situation is quite different near the optimal doping. At the
average doping, DDW is weak or absent. But with increasing $N$,
charge redistribution introduces hole-poor inner layers with DDW
order, counteracting the effect of inter-layer tunnelling and
eventually suppressing $T_c$, as in Fig.3(b).

\begin{figure}
\includegraphics[width=8.5cm,height=6cm]{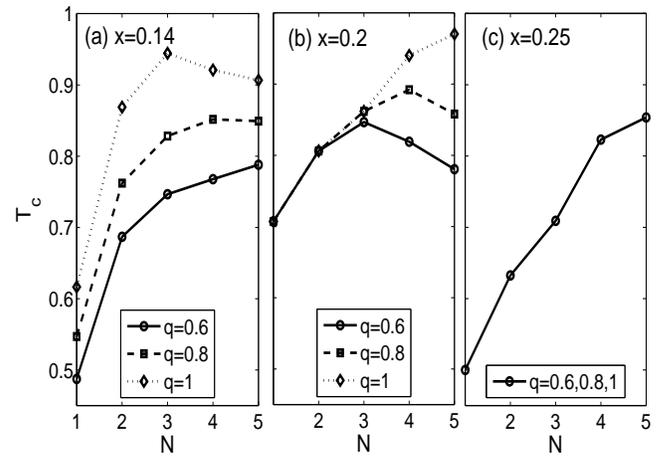}
\caption{The layer-number $N$ dependence of the superconducting
transition temperature $T_c$. (a) $x=0.14$, (b) $x=0.20$, (c)
$x=0.25$.}
\end{figure}

To summarize, we proposed a finite temperature Landau theory that
describes competing orders and interlayer tunneling in cuprate
superconductors as an important extension to a corresponding
theory at zero temperature.\cite{ch} For given parameters, our
theory determines $T_c$ without any ambiguity. In mono- and
double-layer systems we discuss the relation between zero
temperature order parameter and the associated transition
temperature in the presence of competing orders, and discuss the
puzzling experimental fact that the pseudo-gap temperature is much
higher than the corresponding energy scale near optimum doping.
Applying the theory to multi-layer systems, we calculate the
layer-number dependence of $T_c$. In a reasonable parameter space
the result turns out to be in agreement with experiments.

We should emphasize that just as in Ref.\cite{ch} DDW is used as
an example of the competing order to superconductivity. The
discussion in this paper holds for any other forms of competing
orders. However, DDW has the special property that it yields a
band-gap with d-wave symmetry in the momentum space,\cite{ddw} in
accordance with angle-resolved-photoemission
measurement.\cite{arpes} While this seems to be appealing, other
consequences that follow the DDW picture are still in debates. For
example, the temperature derivative of the superfluid density at
low temperatures was claimed to be insensitive to the doping deep
in the under-doped limit of YBCO thin films,\cite{lemberger} but
another group reported that in YBCO crystals it tends to diverge
with under-doping.\cite{diverge} While DDW picture is consistent
with the latter, it can not account for the former
results.\cite{wang} The fate of the DDW picture depends on further
experimental clarification.

\acknowledgments{This work was supported by NSFC 10204011,
10325416, 10021001, and the Fok Ying Tung Education Foundation
No.91009.}


\begin{references}
\bibitem{anderson} P. W. Anderson, Science {\bf 235}, 1196 (1987)
\bibitem{ch} S. Chakravarty, H. Y. Kee, and K. V\"{o}lker, Nature {\bf 428}, 53 (2004).
\bibitem{TcN} B. A. Scott, E. Y. Suard, C. C. Tsuei, D. G. Mitzi, T. R. McGuire, B.-H. Chen,
and D. Walker, Physica C {\bf 230}, 239 (1994); I. G. Kuzemskaya,
A. L. Kuzemsky, and A. A. Cheglokov, J. Low Temp. Phys. {\bf 118},
147 (2000).
\bibitem{ddw} S. Chakravarty, R. B. Laughlin, D. K. Morr and C.
Nayak, Phys. Rev. B. {\bf 63}, 094503 (2001).
\bibitem{interlayer} J. M. Wheatley, T. C. Hsu, P. W. Anderson,
Nature {\bf 333}, 121 (1988); S. Chakravarty, A. Sudbo, P. W.
Anderson, S. Strong, Science {\bf 261}, 337 (1993); P. W.
Anderson, Science {\bf 268}, 1154 (1995)
\bibitem{loram} J. L. Tallon, G. V. M. Williams, and J. W. Loram, Physica C {\bf 338},
9 (2000).
\bibitem{phase} T. A. Zaleski and T. K. Kope\'{c},
cond-mat/0406142.
\bibitem{charge} H. Kotegawa, Y. Tokunaga, K. Ishida, G.-q. Zheng, Y. Kitaoka, H. Kito,
A. Iyo, K. Tokiwa, T. Watanabe, and H. Ihara, Phys. Rev. B {\bf
64}, 064515 (2001); H. Kotegawa, Y. Tokunaga, K. Ishida, G.-q.
Zheng, Y. Kitaoka, K. Asayama, H. Kito, A. Iyo, H. Ihara, K.
Tanaka, K. Tokiwa, and T. Watanabe, J. Phys. Chem. Solids. {\bf
62}, 171(2001); .
\bibitem{arpes} H. Ding, T. Yokoya, J. C. Campuzano, T. Takahashi, M. Randeria, M. R. Norman,
T. Mochiku, K. Kadowaki, and J. Giapintzakis, Nature {\bf 382}, 51
(1996); A. G. Loeser, Z.-X. Shen, D. S. Dessau, D. S. Marshall, C.
H. Park, P. Fournier, and A. Kapitulnik, Science {\bf 273}, 325
(1996); D. S. Marshall, D. S. Dessau, A. G. Loeser, C-H. Park, A.
Y. Matsuura, J. N. Eckstein, I. Bozovic, P. Fournier, A.
Kapitulnik, W. E. Spicer, and Z.-X. Shen, Phys. Rev. Lett. {\bf
76}, 4841 (1996).
\bibitem{lemberger} A. Hosseini, D.M. Broun,D. E. Sheehy,T. P. Davis, M. Franz,W. N. Hardy,
Ruixing Liang, and D. A. Bonn, Phys. Rev. Lett. {\bf 93}, 107003
(2004); D. A. Bonn, S. Kamal, A. Bonakdarpour, R. X. Liang, W. N.
Hardy, C. C. Homes, D. N. Basov, T. Timusk, Czech. J. Phys. {\bf
46}, 3195 (1996); B. R. Boyce, J. A. Skinta, and T. R. Lemberger,
Physica C {\bf 341}, 561 (1997).
\bibitem{diverge} M. R. Trunin, Y. A. Nefyodov, and A. F.
Shevchun, Phys. Rev. Lett. {\bf 92}, 067006 (2004).
\bibitem{wang} Qiang-Hua Wang, Jung Hoon Han, and Dung-Hai Lee,
Phys. Rev. Lett. {\bf 87}, 077004 (2001).
\end{references}
\end{document}